# Generalized Dynamics for Second Kind of Soft-Matter Quasicrystals

Tian-You Fan

School of Physics, Beijing Institute of Technology, Beijing 100081, China
tyfan2013@163.com

**Abstract** The soft-matter quasicrystals observed so far are two-dimensional quasicrystals. A part of equation systems of generalized hydrodynamics, or generalized dynamics for simplicity, for soft-matter quasicrystals are reported in Ref [1], in which the 12-fold and possible 5-, 10- and 8-fold symmetry soft-matter quasicrystals are identified as the first kind of soft-matter quasicrystals, while the 18-fold and possible 7-, 9- and 14-fold symmetry ones belong to the second kind of soft-matter quasicrystals，based on the point of view of symmetry. This paper reports the equations of generalized dynamics of second kind of soft-matter quasicrystals. For simplicity the discussion is only limited to the two-dimensional case.

**Keywords**： soft matter; second kind 2D quasicrystals; 6D embedding space; generalized dynamics; equation of state

**Introduction**

Ref [1] reported quasicrystals with 12- and 18-fold symmetry quasicrystals in soft matter and possible 5-, 8- and 10-fold symmetry quasicrystals in the matter. It is well-known the 12-fold symmetry quasicrystals are observed in many categories of soft matter, and 18-fold symmetry quasicrystals in colloids were observed [2] too. From the angle of symmetry, the 18-fold symmetry quasicrystals are different from those of 12- and 5-, 8- and 10-fold symmetry quasicrystals. There are 7-, 9- and 14-fold symmetry quasicrystals are similar to those of 18-fold symmetry, but have never been observed in solid as well as in soft matter yet. According to the symmetry theory the 5-, 8-, 10- and 12-fold symmetry quasicrystals belong to the first kind of two-dimensional quasicrystals, while 7-, 9- , 14- and 18-fold symmetry quasicrystals are classified as the second kind of two-dimensional quasicrystals. In this letter we will report the equation systems of generalized dynamics of the 7-, 9- and 14-fold symmetry quasicrystals, because the equations of 18-fold symmetry quasicrystals in soft matter are reported already in Ref [1]. The key point is the six-dimensional space of Hu et al [3], in which those authors developed the group theory and group representation theory, and predicted the existence of the second two-dimensional quasicrystals before 17 years in the experimental observation; this is an important contribution of Chinese scientists in the field of quasicrystals. Hu et al in Ref [3] also determined the all nonzero independent elastic constants of the second kind of two-dimensional quasicrystals by using the theory of group representation, which is important contribution for studying the dynamics of solid and soft-matter quasicrystals too. With the basic contribution of Hu et al, and introduce the fluid phonon and equation of state of soft matter similarly in Ref [1] and introducing the Landau-Anderson [4,5] symmetry breaking and elementary excitation principle, we set up the generalized dynamics of the soft-matter quasicrystals.

**1. Six-dimensional embedding space and elementary excitations of the second kind of soft-matter quasicrystals**

The two-dimensional solid quasicrystals observed thus far are of the first kind, i.e., they present 5-, 8-,10- and 12-fold symmetries. In symmetry theory, 5-, 8-, 10- and 12-fold symmetry quasicrystals—either in solid or in soft matter—belong to a similar quasiperiodic structure and present: (1) One needs a set of four rationally independent reciprocal basis vectors to index the diffraction pattern with integers; (2) The basis vectors can be considered as a projection from a 4-dimensional embedding space ($V$) upon a 2-dimensional physical space ($V_E$); (3) The space $V$ is the direct sum of $V_E$ and $V_I$ where $V_I$ is the orthogonal complementary space; and (4) Four hydrodynamic degrees of freedom in phases can be parameterized by the two-dimensional vector field. One of these is the phonon field (denoted by **u**), and the other is the phason field (denoted by **w**). Quasicrystals exhibiting these characteristics are first kind two-dimensional quasicrystals.

Unlike the first kind of two-dimensional quasicrystals, the second type of two-dimensional quasicrystals with 7-, 9-, 14- and 18-fold symmetries present very different characteristics of quasiperiodic symmetry and are second kind two-dimensional quasicrystals. To describe the physical properties of second kind quasicrystals, Hu et al analyzed one have to introduce a so-called concept of six-dimensional embedding space [3], which leads to phonon field **u**, the first phason field **v**, and the second phason field **w**, respectively, and will be further discussed.

So far, all observed solid two-dimensional quasicrystals are first kind two-dimensional quasicrystals, while those discovered in soft matter consisted of both kinds of two-dimensional quasicrystals. Thus, soft-matter quasicrystals can be found in more quasicrystal systems, and present more development potential.

The distinction of 7-, 9-, 14- and 18-fold quasicrystals to those of 5-, 8-, 10- and 12-fold symmetries leads to the appearance of the so-called six-dimensional embedding space, leads to the extension [3] of the expansion of the Landau-Anderson expansion [4,5], suggesting second kind phason elementary excitation (quasiparticle), and promotes the development of condensed matter physics. The discovery of soft-matter quasicrystals of 12- and 18-fold symmetries creates in-depth insights into quasicrystal study.

These 7-, 9- ,14- and 18-fold symmetry quasicrystals have not been observed in solid and 5-, 8-, 10-, 7-, 9- and 14-fold symmetry quasicrystals have not yet been discovered in soft matter. The study of these classes of quasicrystals may be significant from the angle of symmetry and is beneficial for exploring structures and physical properties (including mechanical properties); and for the setting up the generalized dynamics of the matter. Due to space limitations, the discussion in this paper is limited to soft-matter quasicrystals.

It has been previously mentioned that there are principle distinctions between the first and second kinds of two-dimensional quasicrystals. This led to the extension of the concepts on parallel space $E_\parallel^3$ and perpendicular space $E_\perp^3$ used in conventional quasicrystal study must be replaced by parallel space $E_\parallel^2$, and the first and second perpendicular spaces $E_{\perp 1}^2$ and $E_{\perp 2}^2$ formed the six-dimensional embedding space for the second kind two-dimensional quasicrystal study [3], i.e.,

$$E^6 = E_\parallel^2 \oplus E_{\perp 1}^2 \oplus E_{\perp 2}^2 \tag{1}$$

Based on the above concept, the well-known Landau-Anderson expansion [3.4],

$$\rho(\mathbf{r}) = \sum_{\mathbf{G} \in L_R} \rho_\mathbf{G} \exp\{i\mathbf{G} \cdot \mathbf{r}\} = \sum_{\mathbf{G} \in L_R} |\rho_\mathbf{G}| \exp\{-i\Phi_\mathbf{G} + i\mathbf{G} \cdot \mathbf{r}\} \quad (2)$$

needs to be extended for the phase (phase angle) in Equation (2) by

$$\Phi_n = \mathbf{G}_n^{\parallel} \cdot \mathbf{u} + \mathbf{G}_n^{\perp 1} \cdot \mathbf{v} + \mathbf{G}_n^{\perp 2} \cdot \mathbf{w} \quad (3)$$

where $\mathbf{G}_n^{\parallel}$ expresses reciprocal lattice vector in parallel space; $\mathbf{G}_n^{\perp 1}$ and $\mathbf{G}_n^{\perp 2}$ are the reciprocal lattice vectors in the first and second perpendicular spaces; $\mathbf{u}$ is the phonon displacement field; $\mathbf{v}$ and $\mathbf{w}$ are the first and second phason displacement fields in the first and second perpendicular spaces, respectively.

Although in the sense of elementary excitations, the fields $\mathbf{u}$, $\mathbf{v}$ and $\mathbf{w}$ are in similar places, in generalized dynamics $\mathbf{u}$ represents wave propagation, and $\mathbf{v}$ and $\mathbf{w}$ represent diffusion, which are quite different to each other in physical meaning.

The six-dimensional embedding space in Equation (1) and the Landau-Anderson expansions in Equations (2) and (3) is the basis of the following discussion. The expansions in Equations (2) and (3) contain many fruitful insights. First, they belong to an application of the Landau principle of breaking symmetry and elementary excitation (quasiparticle). Essentially, the principle describes the quantum mechanics on the collective excitation of massive atoms, which is a result of quantization. It holds the key to the phase transition of a higher-ordered phase to a lower-ordered phase, and leads to symmetry breaking and the appearance of a new order parameter. The order parameter here is $\rho_\mathbf{G}$ in Equation (2), and it is connected closely to the phase (phase angle). The concept on phase is important like that of quantization in modern physics. It is well-known, quantization, symmetry and phase constitute three main themes of modern physics starting from Schrödinger and Heisenberg, and the study of quasicrystals are closed of connection with the three main themes, this shows in-depth insight of the science of quasicrystals.

## 2. Equations of Generalized Dynamics of Soft-Matter Quasicrystals with 18-Fold Symmetry

The equations have been reported in the Ref [1], so it is omitted here.

## 3. Generalized Dynamics of Soft-Matter Quasicrystals with 7-FoldSymmetry

In the previous section, the elementary excitations of phonons, first and second phasons in quasicrystals of 18-fold symmetry were introduced, the detail please refer to Ref [1]. The 7-fold symmetry quasicrystals are significant in the second kind of two-dimensional quasicrystals, of which features present different behaviour from 18-fold symmetry quasicrystals, although they both belong to the second kind two-dimensional quasicrystals.

According to the theory of group representation, deformation energy density and the relevant constitutive law for the 7-fold symmetry quasicrystals in soft matter as follows:

$$\left.\begin{aligned}
\sigma_{xx} &= (L+2M)\varepsilon_{xx} + L\varepsilon_{yy} + R(w_{xx}+w_{yy}) \\
\sigma_{yy} &= L\varepsilon_{xx} + (L+2M)\varepsilon_{yy} - R(w_{xx}+w_{yy}) \\
\sigma_{xy} &= \sigma_{yx} = 2M\varepsilon_{xy} + R(w_{yx}-w_{xy}) \\
\tau_{xx} &= T_1 v_{xx} + T_2 v_{yy} + G\left(w_{xx}-w_{yy}\right) \\
\tau_{yy} &= T_2 v_{xx} + T_1 v_{yy} - G\left(w_{xx}-w_{yy}\right) \\
\tau_{xy} &= T_1 v_{xy} - T_2 v_{yx} + G\left(w_{yx}+w_{xy}\right) \\
\tau_{yx} &= -T_2 v_{xy} + T_1 v_{yx} + G\left(w_{yx}+w_{xy}\right) \\
H_{xx} &= K_1 w_{xx} + K_2 w_{yy} + R(\varepsilon_{xx}-\varepsilon_{yy}) + G\left(v_{xx}-v_{yy}\right) \\
H_{yy} &= K_2 w_{xx} + K_1 w_{yy} + R(\varepsilon_{xx}-\varepsilon_{yy}) - G\left(v_{xx}-v_{yy}\right) \\
H_{xy} &= K_1 w_{xy} - K_2 w_{yx} - 2R\varepsilon_{xy} + G\left(v_{xy}+v_{yx}\right) \\
H_{yx} &= K_1 w_{yx} - K_2 w_{xy} + 2R\varepsilon_{xy} + G\left(v_{xy}+v_{yx}\right) \\
p_{xx} &= -p + 2\eta(\dot{\xi}_{xx} - \frac{1}{3}\dot{\xi}_{kk}) \\
p_{yy} &= -p + 2\eta(\dot{\xi}_{yy} - \frac{1}{3}\dot{\xi}_{kk}) \\
p_{xy} &= p_{yx} = 2\eta\dot{\xi}_{xy} \\
\dot{\xi}_{kk} &= \dot{\xi}_{xx} + \dot{\xi}_{yy}
\end{aligned}\right\} \quad (4)$$

and based on a generalized Langevin equation and Poisson bracket method, with the addition of an equation of state, the dynamic equations for the present quasicrystal system are obtained as follows:

$$\frac{\partial \rho}{\partial t} + \nabla \cdot (\rho \mathbf{V}) = 0$$

$$\frac{\partial (\rho V_x)}{\partial t} + \frac{\partial (V_x \rho V_x)}{\partial x} + \frac{\partial (V_y \rho V_x)}{\partial y} = -\frac{\partial p}{\partial x} + \eta \nabla^2 V_x + \frac{1}{3}\eta \frac{\partial}{\partial x} \nabla \cdot \mathbf{V} + M\nabla^2 u_x + (L+M-B)\frac{\partial}{\partial x}\nabla \cdot \mathbf{u}$$
$$-(A-B)\frac{1}{\rho_0}\frac{\partial \delta \rho}{\partial x}$$

$$\frac{\partial (\rho V_y)}{\partial t} + \frac{\partial (V_x \rho V_y)}{\partial x} + \frac{\partial (V_y \rho V_y)}{\partial y} = -\frac{\partial p}{\partial y} + \eta \nabla^2 V_y + \frac{1}{3}\eta \frac{\partial}{\partial y} \nabla \cdot \mathbf{V} + M\nabla^2 u_y + (L+M-B)\frac{\partial}{\partial y}\nabla \cdot \mathbf{u}$$
$$-(A-B)\frac{1}{\rho_0}\frac{\partial \delta \rho}{\partial y}$$

$$\frac{\partial u_x}{\partial t} + V_x \frac{\partial u_x}{\partial x} + V_y \frac{\partial u_x}{\partial y} = V_x + \Gamma_{\mathbf{u}}\left[M\nabla^2 u_x + (L+M)\frac{\partial}{\partial x}\nabla \cdot \mathbf{u} + R\left(\frac{\partial^2 w_x}{\partial x^2} + 2\frac{\partial^2 w_y}{\partial x \partial y} - \frac{\partial^2 w_x}{\partial y^2}\right)\right]$$

$$\frac{\partial u_y}{\partial t} + V_x \frac{\partial u_y}{\partial x} + V_y \frac{\partial u_y}{\partial y} = V_y + \Gamma_{\mathbf{u}}\left[M\nabla^2 u_y + (L+M)\frac{\partial}{\partial y}\nabla \cdot \mathbf{u} + R\left(\frac{\partial^2 w_y}{\partial x^2} - 2\frac{\partial^2 w_x}{\partial x \partial y} - \frac{\partial^2 w_y}{\partial y^2}\right)\right]$$

$$\frac{\partial v_x}{\partial t} + V_x \frac{\partial v_x}{\partial x} + V_y \frac{\partial v_x}{\partial y} = \Gamma_v \left[T_1 \nabla^2 v_x + G \nabla^2 w_x\right]$$

$$\frac{\partial v_y}{\partial t} + V_x \frac{\partial v_y}{\partial x} + V_y \frac{\partial v_y}{\partial y} = \Gamma_v \left[T_1 \nabla^2 v_y + G \nabla^2 w_y\right]$$

$$\frac{\partial w_x}{\partial t} + V_x \frac{\partial w_x}{\partial x} + V_y \frac{\partial w_x}{\partial y} = \Gamma_{\mathbf{w}}\left[K_1 \nabla^2 w_x + R\left(\frac{\partial^2 u_x}{\partial x^2} - 2\frac{\partial^2 u_y}{\partial x \partial y} - \frac{\partial^2 u_x}{\partial y^2}\right) + G\nabla^2 v_x\right]$$

$$\frac{\partial w_y}{\partial t} + V_x \frac{\partial w_y}{\partial x} + V_y \frac{\partial w_y}{\partial y} = \Gamma_{\mathbf{w}}\left[K_1 \nabla^2 w_y + R\left(\frac{\partial^2 u_y}{\partial x^2} + 2\frac{\partial^2 u_x}{\partial x \partial y} - \frac{\partial^2 u_y}{\partial y^2}\right) + G\nabla^2 v_y\right]$$

$$p = f(\rho) = 3\frac{k_B T}{l^3 \rho_0^3}\left(\rho_0^2 \rho + \rho_0 \rho^2 + \rho^3\right)$$

(5)

where $\nabla^2 = \frac{\partial^2}{\partial x^2} + \frac{\partial^2}{\partial y^2}$, $\nabla = \mathbf{i}\frac{\partial}{\partial x} + \mathbf{j}\frac{\partial}{\partial y}$, $\mathbf{V} = \mathbf{i}V_x + \mathbf{j}V_y$, $\mathbf{u} = \mathbf{i}u_x + \mathbf{j}u_y$, and $L = C_{12}$, $M = (C_{11} - C_{12})/2$ are phonon elastic constants; $T_1, K_1$ are the elastic constants of the first and second phasons; $R, G$ are the coupling elastic constants between the phonons and second phasons and between the first and second phasons; $\eta$ is the fluid viscosity (for simplicity, only the scalar version of $\eta_{ijkl}$ was considered ); $\Gamma_u, \Gamma_v$ and $\Gamma_w$ are the phonon, first phason and second phason dissipation coefficients, respectively; and $A$, $B$ are the material constants due to the variation of mass density.

Equation (5) consists of 10 field variables, i.e., phonon field $\mathbf{u} = (u_x, u_y)$, first phason field $\mathbf{v} = (v_x, v_y)$ and second phason field $\mathbf{w} = (w_x, w_y)$, fluid velocity field $\mathbf{V} = (V_x, V_y)$, mass density $\rho$ and fluid pressure $p$, respectively. The amount of the equations is also 10: (5a) is the mass conservation equation or the continuum equation;(5b) and (5c) are the momentum conservation equations or the generalized Navier-Stokes equations;(5d) and (5e) are the equations of motion of phonons due to symmetry breaking; (5f) and (5g) are the first phason dissipation equations;(5h) and (5i) are the second phason dissipation equations; and (5j) is the equation of state. If there is no the equation of state, the equation system is not closed, and the importance of the equation is evident, as given by Ref [1].The equations reveal the nature of wave propagation of fields **u** and **V**, and the nature of diffusion of fields **v** and **w** from the view of hydrodynamics. In the present case, Equation (5) is consistent mathematically and solvable.

## 4. Generalized Dynamics of Soft-Matter Quasicrystals with 14-Fold Symmetry

Quasicrystals of 14-fold symmetry in soft matter are also very interesting，and consist of 4 kinds of elementary excitations where phonons and second phasons are coupled, as that are first and second phasons.

With the theory of group representation we can find the constitutive law of 14-fold symmetry quasicrystals in soft matter such as

$$\left.\begin{aligned}
\sigma_{xx} &= (L+2M)\varepsilon_{xx} + L\varepsilon_{yy} + R(w_{xx}+w_{yy}) \\
\sigma_{yy} &= L\varepsilon_{xx} + (L+2M)\varepsilon_{yy} - R(w_{xx}+w_{yy}) \\
\sigma_{xy} &= \sigma_{yx} = 2M\varepsilon_{xy} + R(w_{yx}-w_{xy}) \\
\tau_{xx} &= T_1 v_{xx} + T_2 v_{yy} + G(w_{xx}-w_{yy}) \\
\tau_{yy} &= T_2 v_{xx} + T_1 v_{yy} + G(w_{xx}-w_{yy}) \\
\tau_{xy} &= T_1 v_{xy} - T_2 v_{yx} - G(w_{yx}+w_{xy}) \\
\tau_{yx} &= -T_2 v_{xy} + T_1 v_{yx} + G(w_{yx}+w_{xy}) \\
H_{xx} &= K_1 w_{xx} + K_2 w_{yy} + R(\varepsilon_{xx}-\varepsilon_{yy}) + G(v_{xx}+v_{yy}) \\
H_{yy} &= K_2 w_{xx} + K_1 w_{yy} + R(\varepsilon_{xx}-\varepsilon_{yy}) - G(v_{xx}+v_{yy}) \\
H_{xy} &= K_1 w_{xy} - K_2 w_{yx} - 2R\varepsilon_{xy} + G(v_{yx}-v_{xy}) \\
H_{yx} &= K_1 w_{yx} - K_2 w_{xy} + 2R\varepsilon_{xy} + G(v_{yx}-v_{xy}) \\
p_{xx} &= -p + 2\eta(\dot{\xi}_{xx} - \frac{1}{3}\dot{\xi}_{kk}) \\
p_{yy} &= -p + 2\eta(\dot{\xi}_{yy} - \frac{1}{3}\dot{\xi}_{kk}) \\
p_{xy} &= p_{yx} = 2\eta\dot{\xi}_{xy} \\
\dot{\xi}_{kk} &= \dot{\xi}_{xx} + \dot{\xi}_{yy}
\end{aligned}\right\} \quad (6)$$

so the equation system of dynamics of the quasicrystals can be derived as:

$$\frac{\partial \rho}{\partial t}+\nabla\bullet(\rho\mathbf{V})=0$$

$$\frac{\partial(\rho V_x)}{\partial t}+\frac{\partial(V_x\rho V_x)}{\partial x}+\frac{\partial(V_y\rho V_x)}{\partial y}=-\frac{\partial p}{\partial x}+\eta\nabla^2 V_x+\frac{1}{3}\eta\frac{\partial}{\partial x}\nabla\bullet\mathbf{V}+M\nabla^2 u_x+(L+M-B)\frac{\partial}{\partial x}\nabla\bullet\mathbf{u}$$
$$-(A-B)\frac{1}{\rho_0}\frac{\partial\delta\rho}{\partial x}$$

$$\frac{\partial(\rho V_y)}{\partial t}+\frac{\partial(V_x\rho V_y)}{\partial x}+\frac{\partial(V_y\rho V_y)}{\partial y}=-\frac{\partial p}{\partial y}+\eta\nabla^2 V_y+\frac{1}{3}\eta\frac{\partial}{\partial y}\nabla\bullet\mathbf{V}+M\nabla^2 u_y+(L+M-B)\frac{\partial}{\partial y}\nabla\bullet\mathbf{u}$$
$$-(A-B)\frac{1}{\rho_0}\frac{\partial\delta\rho}{\partial y}$$

$$\frac{\partial u_x}{\partial t}+V_x\frac{\partial u_x}{\partial x}+V_y\frac{\partial u_x}{\partial y}=V_x+\Gamma_\mathbf{u}\left[M\nabla^2 u_x+(L+M)\frac{\partial}{\partial x}\nabla\bullet\mathbf{u}+R\left(\frac{\partial^2 w_x}{\partial x^2}+2\frac{\partial^2 w_y}{\partial x\partial y}-\frac{\partial^2 w_x}{\partial y^2}\right)\right]$$

$$\frac{\partial u_y}{\partial t}+V_x\frac{\partial u_y}{\partial x}+V_y\frac{\partial u_y}{\partial y}=V_y+\Gamma_\mathbf{u}\left[M\nabla^2 u_y+(L+M)\frac{\partial}{\partial y}\nabla\bullet\mathbf{u}+R\left(\frac{\partial^2 w_y}{\partial x^2}-2\frac{\partial^2 w_x}{\partial x\partial y}-\frac{\partial^2 w_y}{\partial y^2}\right)\right]$$

$$\frac{\partial v_x}{\partial t}+V_x\frac{\partial v_x}{\partial x}+V_y\frac{\partial v_x}{\partial y}=\Gamma_v\left[T_1\nabla^2 v_x+G\left(\frac{\partial^2 w_x}{\partial x^2}-\frac{\partial^2 w_x}{\partial y^2}\right)-2G\frac{\partial^2 w_y}{\partial x\partial y}\right]$$

$$\frac{\partial v_y}{\partial t}+V_x\frac{\partial v_y}{\partial x}+V_y\frac{\partial v_y}{\partial y}=\Gamma_v\left[T_1\nabla^2 v_y+2G\frac{\partial^2 w_x}{\partial x\partial y}+G\left(\frac{\partial^2 w_y}{\partial x^2}-\frac{\partial^2 w_y}{\partial y^2}\right)\right]$$

$$\frac{\partial w_x}{\partial t}+V_x\frac{\partial w_x}{\partial x}+V_y\frac{\partial w_x}{\partial y}=\Gamma_\mathbf{w}\left[K_1\nabla^2 w_x+R\left(\frac{\partial^2 u_x}{\partial x^2}-2\frac{\partial^2 u_y}{\partial x\partial y}-\frac{\partial^2 u_x}{\partial y^2}\right)+G\left(\frac{\partial^2 v_x}{\partial x^2}-\frac{\partial^2 v_x}{\partial y^2}\right)+2G\frac{\partial^2 v_y}{\partial x\partial y}\right]$$

$$\frac{\partial w_y}{\partial t}+V_x\frac{\partial w_y}{\partial x}+V_y\frac{\partial w_y}{\partial y}=\Gamma_\mathbf{w}\left[K_1\nabla^2 w_y+R\left(\frac{\partial^2 u_y}{\partial x^2}+2\frac{\partial^2 u_x}{\partial x\partial y}-\frac{\partial^2 u_y}{\partial y^2}\right)-2G\frac{\partial^2 v_x}{\partial x\partial y}-G\left(\frac{\partial^2 v_y}{\partial x^2}-\frac{\partial^2 v_y}{\partial y^2}\right)\right]$$

$$p=f(\rho)=3\frac{k_B T}{l^3\rho_0^3}\left(\rho_0^2\rho+\rho_0\rho^2+\rho^3\right)$$

(7)

where $\nabla^2=\frac{\partial^2}{\partial x^2}+\frac{\partial^2}{\partial y^2}$, $\nabla=\mathbf{i}\frac{\partial}{\partial x}+\mathbf{j}\frac{\partial}{\partial y}$, $\mathbf{V}=\mathbf{i}V_x+\mathbf{j}V_y, \mathbf{u}=\mathbf{i}u_x+\mathbf{j}u_y$, and $L=C_{12}$, $M=(C_{11}-C_{12})/2$ are phonon elastic constants; $T_1, K_1$ are the elastic constants of the first and second phasons; $R, G$ are the coupling elastic constants of the phonons and second phasons, and first and second phasons (in the case there is no coupling between the phonon and first phason, i.e., $r_{ijkl}=0$); $\eta$ is the fluid viscosity (for simplicity, only the scalar version of $\eta_{ijkl}$ was considered); $\Gamma_u, \Gamma_v$ and $\Gamma_w$ are the phonon, first phason and second phason dissipation coefficients, respectively; and $A$, $B$ are the material constants due to the variation of mass density.

Equation (7) contains 10 field variables, i.e., phonon field $\mathbf{u}=(u_x,u_y)$, first phason field $\mathbf{v}=(v_x,v_y)$ and second phason field $\mathbf{w}=(w_x,w_y)$, fluid velocity field $\mathbf{V}=(V_x,V_y)$, mass density $\rho$; and fluid pressure $p$, respectively. The amount of the equations is also 10: (7a) is the mass conservation equation or the continuum equation; (7b) and (7c) are momentum conservation equations or the generalized Navier-Stokes equations; (7d) and (7e)are the equations of motion of phonons due to symmetry breaking;(7f) and (7g) are the first phason dissipation equations; (7h) and (7i)are the second phason dissipation equations; and(7j) is the equation of state. If there is no the equation of state, the equation system is not closed, and the importance of the equation is evident, which is given by Ref [1].The equations reveal the nature of wave propagation of fields **u** and **V** , and the nature of diffusion of field **v** and **w** from the view of hydrodynamics. At present, Equation (7) is mathematically consistent and is solvable.

## 5. Generalized Dynamics of Soft-Matter Quasicrystals with 9-Fold Symmetry

Quasicrystals of 9-fold symmetry in soft matter are different from quasicrystals of 7- and 14-fold, and are very close to those of 18-fold symmetry. Quasicrystals of 9-fold symmetry consist of four kinds of elementary excitations, where the phonons and phasons are decoupled, but the first and second phasons are coupled according to the theory of group representation, by using the theory, we can obtain the constitutive equations for the quasicrystals

$$\left.\begin{aligned}
\sigma_{xx} &= (L+2M)\varepsilon_{xx} + L\varepsilon_{yy} \\
\sigma_{yy} &= L\varepsilon_{xx} + (L+2M)\varepsilon_{yy} \\
\sigma_{xy} &= \sigma_{yx} = 2M\varepsilon_{xy} \\
\tau_{xx} &= T_1 v_{xx} + T_2 v_{yy} + G(w_{xx} + w_{yy}) \\
\tau_{yy} &= T_2 v_{xx} + T_1 v_{yy} - G(w_{xx} + w_{yy}) \\
\tau_{xy} &= T_1 v_{xy} - T_2 v_{yx} + G(w_{yx} - w_{xy}) \\
\tau_{yx} &= -T_2 v_{xy} + T_1 v_{yx} + G(w_{yx} - w_{xy}) \\
H_{xx} &= K_1 w_{xx} + K_2 w_{yy} + G(v_{xx} - v_{yy}) \\
H_{yy} &= K_2 w_{xx} + K_1 w_{yy} + G(v_{xx} - v_{yy}) \\
H_{xy} &= K_1 w_{xy} - K_2 w_{yx} - G(v_{xy} + v_{yx}) \\
H_{yx} &= K_1 w_{yx} - K_2 w_{xy} + G(v_{xy} + v_{yx}) \\
p_{xx} &= -p + 2\eta(\dot{\xi}_{xx} - \frac{1}{3}\dot{\xi}_{kk}) \\
p_{yy} &= -p + 2\eta(\dot{\xi}_{yy} - \frac{1}{3}\dot{\xi}_{kk}) \\
p_{xy} &= p_{yx} = 2\eta\dot{\xi}_{xy} \\
\dot{\xi}_{kk} &= \dot{\xi}_{xx} + \dot{\xi}_{yy}
\end{aligned}\right\} (8)$$

so the dynamic equation system of 9-fold symmetry quasicrystals in soft-matter by similar derivation to those given in previous sections is as follows:

$$\left.\begin{aligned}
&\frac{\partial \rho}{\partial t}+\nabla \cdot(\rho \mathbf{V})=0 \\
&\frac{\partial(\rho V_x)}{\partial t}+\frac{\partial(V_x \rho V_x)}{\partial x}+\frac{\partial(V_y \rho V_x)}{\partial y}=-\frac{\partial p}{\partial x}+\eta \nabla^2 V_x+\frac{1}{3}\eta\frac{\partial}{\partial x}\nabla \cdot \mathbf{V}+M\nabla^2 u_x+(L+M-B)\frac{\partial}{\partial x}\nabla \cdot \mathbf{u}\\
&-(A-B)\frac{1}{\rho_0}\frac{\partial \delta\rho}{\partial x}\\
&\frac{\partial(\rho V_y)}{\partial t}+\frac{\partial(V_x \rho V_y)}{\partial x}+\frac{\partial(V_y \rho V_y)}{\partial y}=-\frac{\partial p}{\partial y}+\eta \nabla^2 V_y+\frac{1}{3}\eta\frac{\partial}{\partial y}\nabla \cdot \mathbf{V}+M\nabla^2 u_y+(L+M-B)\frac{\partial}{\partial y}\nabla \cdot \mathbf{u}\\
&-(A-B)\frac{1}{\rho_0}\frac{\partial \delta\rho}{\partial y}\\
&\frac{\partial u_x}{\partial t}+V_x\frac{\partial u_x}{\partial x}+V_y\frac{\partial u_x}{\partial y}=V_x+\Gamma_\mathbf{u}\left[M\nabla^2 u_x+(L+M)\frac{\partial}{\partial x}\nabla \cdot \mathbf{u}\right]\\
&\frac{\partial u_y}{\partial t}+V_x\frac{\partial u_y}{\partial x}+V_y\frac{\partial u_y}{\partial y}=V_y+\Gamma_\mathbf{u}\left[M\nabla^2 u_y+(L+M)\frac{\partial}{\partial y}\nabla \cdot \mathbf{u}\right]\\
&\frac{\partial v_x}{\partial t}+V_x\frac{\partial v_x}{\partial x}+V_y\frac{\partial v_x}{\partial y}=\Gamma_v\left[T_1\nabla^2 v_x+G\left(\frac{\partial^2 w_x}{\partial x^2}-\frac{\partial^2 w_x}{\partial y^2}\right)-2G\frac{\partial^2 w_y}{\partial x\partial y}\right]\\
&\frac{\partial v_y}{\partial t}+V_x\frac{\partial v_y}{\partial x}+V_y\frac{\partial v_y}{\partial y}=\Gamma_v\left[T_1\nabla^2 v_y+2G\frac{\partial^2 w_x}{\partial x\partial y}+G\left(\frac{\partial^2 w_y}{\partial x^2}-\frac{\partial^2 w_y}{\partial y^2}\right)\right]\\
&\frac{\partial w_x}{\partial t}+V_x\frac{\partial w_x}{\partial x}+V_y\frac{\partial w_x}{\partial y}=\Gamma_\mathbf{w}\left[K_1\nabla^2 w_x+G\left(\frac{\partial^2 v_x}{\partial x^2}-\frac{\partial^2 v_x}{\partial y^2}\right)+2G\frac{\partial^2 v_y}{\partial x\partial y}\right]\\
&\frac{\partial w_y}{\partial t}+V_x\frac{\partial w_y}{\partial x}+V_y\frac{\partial w_y}{\partial y}=\Gamma_\mathbf{w}\left[K_1\nabla^2 w_y-2G\frac{\partial^2 v_x}{\partial x\partial y}+G\left(\frac{\partial^2 v_y}{\partial x^2}-\frac{\partial^2 v_y}{\partial y^2}\right)\right]\\
&p=f(\rho)=3\frac{k_B T}{l^3 \rho_0^3}\left(\rho_0^2\rho+\rho_0\rho^2+\rho^3\right)
\end{aligned}\right\} \quad (9)$$

where $\nabla^2=\frac{\partial^2}{\partial x^2}+\frac{\partial^2}{\partial y^2}$, $\nabla=\mathbf{i}\frac{\partial}{\partial x}+\mathbf{j}\frac{\partial}{\partial y}$, $\mathbf{V}=\mathbf{i}V_x+\mathbf{j}V_y$, $\mathbf{u}=\mathbf{i}u_x+\mathbf{j}u_y$, and $L=C_{12}$, $M=(C_{11}-C_{12})/2$ are phonon elastic constants; $T_1, K_1$ are the elastic constants of the first and second phasons; $G$ are the coupling elastic constants between the first and second phasons (in the case $r_{ijkl}=0$, $R_{ijkl}=0$); $\eta$ is the fluid viscosity (for simplicity, only the scalar version of $\eta_{ijkl}$ was considered), $\Gamma_u, \Gamma_v$ and $\Gamma_w$ the phonon, first phason and second phason dissipation coefficients, respectively; and $A$, $B$ are the material constants due to the variation of mass density.

Equation (9) contains 10 field variables, i.e., phonon field $\mathrm{u}=(u_x,u_y)$, first phason field $\mathrm{v}=(v_x,v_y)$ and second phason field $\mathrm{w}=(w_x,w_y)$, fluid velocity field $\mathrm{V}=(V_x,V_y)$, mass density $\rho$ and fluid pressure $p$, respectively. The amount of the equation is also 10: (9a) is the mass conservation equation or continuum equation;(9b) and (9c) are momentum conservation equations or generalized Navier-Stokes equations;(9d) and (9e) are the phonon equations of motion due to symmetry

breaking;(9f) and (9g) are the first phason dissipation equations; (9h) and (9i) are the second phason dissipation equations; and (9j) is the equation of state. If there is no the equation of state, the equation system is not closed, and the importance of the equation is evident and can be found in Ref[1].The equations reveal the nature of wave propagation of fields **u** and **V**, and the nature of diffusion of fields **v** and **w** from the view of hydrodynamics.

At present, Equation set (9) is mathematically consistent and solvable. Equation (9) is closed to that of 18-foldsymmetry quasicrystals in soft-matter given in Ref [1], but there are some small distinctions between them.

### 6. Brief Introduction on Solutions

The Equation systems(5), (7) and (9), are the governing equation systems of generalized dynamics of soft-matter quasicrystals with 7-, 14- and 9-fold symmetries respectively observed possibly in the near future. These are novel nonlinear partial differential equations and provide the basis for studying dynamics of soft-matter quasicrystals. After establishment of these equations, the author and his students develop a solving system for constructing the solutions of some initial-and boundary-value problems which will be reported in other publications.

### 7. Conclusion and Discussion

The equation systems of generalized dynamics of second kind two-dimensional quasicrystals in soft matter were derived.

The solutions of the initial- and boundary-value problems of these nonlinear partial differential equations have provided fruitful results describing matter distribution, deformation and motion of soft-matter quasicrystals, which present great differences in behaviour physically to those of simple fluids and of solid quasicrystals.

In the theoretical system, the Ref [1] is a basis, and also refer to books [6,7].

The dynamics is a heritage and development of the hydrodynamics of solid quasicrystals by Lubensky et al. [8].

**Acknowledgements** The author thanks the National Natural Science Foundation of China for the support through Grant 11272053; thanks also to Professors T C Lubensky at the University of Pennsylvania, USA; Stephen Z D Cheng at the University of Akron, USA; H H Wensink at the Utrecht University in The Netherlands; and Xian-Fang Li at the Central South University in China for beneficial discussions and kind assistance.

**Appendix The differential-variational form of equations of motion of hydrodynamics of soft-matter quasicrystals (it does not include the equation of state)**

The equations of motion of soft-matter quasicrystals are derived by the Poisson bracket method, which is the heritage and development of Lubensky et al [8]for solid quasicrystals, in which the key is the Hamiltonians for individual quasicrystal systems. For the first kind of two-dimensional quasicrystals of soft matter, the energy functional or the Hamiltonians are similar to that given by Lubensky et al for solid quasicrystals in form for the first kind of two-dimensional quasicrystals in soft matter such as:

$$H = H[\Psi(\mathbf{r},t)] = \int \frac{g^2}{2\rho} d^d\mathbf{r} + \int \left[\frac{1}{2}A\left(\frac{\delta\rho}{\rho_0}\right)^2 + B\left(\frac{\delta\rho}{\rho_0}\right)\nabla\cdot\mathbf{u}\right]d^d\mathbf{r} + F_{el}$$

$$= H_{kin} + H_{density} + F_{el}$$

$$\mathbf{g} = \rho\mathbf{V}, \quad F_{el} = F_u + F_w + F_{uw} \tag{A1}$$

where $F_{el}$ denotes the elastic strain energy; and $F_u, F_w, F_{uw}$ represent the strain energies of phonons, phasons and phonon-phason coupling of the matter, respectively:

$$F_u = \int \frac{1}{2}C_{ijkl}\varepsilon_{ij}\varepsilon_{kl} d^d\mathbf{r}$$

$$F_w = \int \frac{1}{2}K_{ijkl}w_{ij}w_{kl} d^d\mathbf{r} \tag{A2}$$

$$F_{uw} = \int \left(R_{ijkl}\varepsilon_{ij}w_{kl} + R_{klij}w_{ij}\varepsilon_{kl}\right) d^d\mathbf{r}$$

Due to the difference of constitutive laws between soft-matter quasicrystals and solid quasicrystals, the results of the Hamiltonians for soft-matter quasicrystals are different from those of solid quasicrystals.

For the second kind of two-dimensional quasicrystals of soft matter, the elastic energy will be

$$F_{el} = F_u + F_v + F_w + F_{uv} + F_{uw} + F_{vw} \tag{A3}$$

where $F_u, F_v, F_w, F_{uv}, F_{uw}, F_{vw}$ represent the strain energies of phonons, first phasons, second phasons, phonon-first phason coupling, phonon-second phason coupling, and first phason-second phason coupling, respectively:

$$F_u = \int \frac{1}{2}C_{ijkl}\varepsilon_{ij}\varepsilon_{kl} d^d\mathbf{r}$$

$$F_v = \int \frac{1}{2}T_{ijkl}v_{ij}v_{kl} d^d\mathbf{r}$$

$$F_w = \int \frac{1}{2}K_{ijkl}w_{ij}w_{kl} d^d\mathbf{r}$$

$$F_{uv} = \int \left(r_{ijkl}\varepsilon_{ij}v_{kl} + r_{klij}v_{ij}\varepsilon_{kl}\right) d^d\mathbf{r} \tag{A4}$$

$$F_{uw} = \int \left(R_{ijkl}\varepsilon_{ij}w_{kl} + R_{klij}w_{ij}\varepsilon_{kl}\right) d^d\mathbf{r}$$

$$F_{vw} = \int \left(G_{ijkl}v_{ij}w_{kl} + G_{klij}w_{ij}v_{kl}\right) d^d\mathbf{r}$$

Therefore, the problem of the second kind of quasicrystals is more complex than that of the first kind.

With the Hamiltonian (by using the Poisson bracket method) and after lengthy derivation we have the equations of motion for the second kind of two-dimensional quasicrystals in soft matter as follows

$$\frac{\partial \rho}{\partial t} + \nabla_k (\rho V_k) = 0$$

$$\frac{\partial g_i(\mathbf{r},t)}{\partial t} = -\nabla_k(\mathbf{r})(V_k g_i) + \nabla_j(\mathbf{r})\left(-p\delta_{ij} + \eta_{ijkl}\nabla_k(\mathbf{r})V_l\right) - \left(\delta_{ij} - \nabla_i u_j\right)\frac{\delta H}{\delta u_j(\mathbf{r},t)} + \left(\nabla_i v_j\right)\frac{\delta H}{\delta v_j(\mathbf{r},t)}$$

$$+ \left(\nabla_i w_j\right)\frac{\delta H}{\delta w_j(\mathbf{r},t)} - \rho\nabla_i(\mathbf{r})\frac{\delta H}{\delta\rho(\mathbf{r},t)}, \quad g_j = \rho V_j$$

$$\frac{\partial u_i(\mathbf{r},t)}{\partial t} = -V_j\nabla_j(\mathbf{r})u_i - \Gamma_u\frac{\delta H}{\delta u_i(\mathbf{r},t)} + V_i$$

$$\frac{\partial v_i(\mathbf{r},t)}{\partial t} = -V_j\nabla_j(\mathbf{r})v_i - \Gamma_v\frac{\delta H}{\delta v_i(\mathbf{r},t)} \tag{A5}$$

$$\frac{\partial w_i(\mathbf{r},t)}{\partial t} = -V_j\nabla_j(\mathbf{r})w_i - \Gamma_w\frac{\delta H}{\delta w_i(\mathbf{r},t)}$$

The equations for the first kind of two-dimensional quasicrystals in soft matter can be obtained by omitting the field variable $v_i$. Of course, these equations of motion do not include the equation of state, which is a thermodynamic result rather than the result of derivation based on the Poisson bracket. The terms of variational form can be reduced to a differential form based on Ref [1].

After further simplifications, from Equation (A5) and including the equation of state, we can obtain the generalized dynamics Equations (5), (7) and (9) for the plane field of individual two-dimensional quasicrystalline systems of second kind in soft matter.

**References**

[1] Fan T Y, Equation Systems of Generalized Hydrodynamics for Soft-Matter Quasicrystals, *Appl. Math. Mech.*,2016,**37**(4),331-347 (in Chinese).
[2]Fischer S, Exner A, Zielske K, Perlich J, Deloudi S, Steuer W, Linder P and Foestor S, Colloidal quasicrystals with 12-fold and 18-fold diffraction symmetry, *Proc Nat Ac Sci*, 2011, **108**, 1810-1814.
[3]Hu C Z, Ding D H, Wang R H and Yang W G, Possible two-dimensional quasicrystals structures with a six-dimensional embedding space, *Phys Rev B*, 1994, **49**(14), 9423-9427.
[4]Landau LD and Lifshitz EM, Statistical Physics, Part 1,1980, Oxford, Butterworth-Heinemann Ltd.
[5]Anderson P W, Basic Notations of Condensed Matter Physics,1984, Menlo Park, Benjamin-Cummings.
[6] Fan T Y, Mathematical Theory of Elasticity of Quasicrystals and Its Applications, 2010, 1st Edition, 2016, 2nd Edition, Beijing, Science Press/Heidelberg, Springer-Verlag.
[7]Fan T Y, Mathematical Theory of Elasticity and Relevant Topics of Solid and Soft-Matter Quasicrystals, Beijing, Beijing Institute Technology Press, 2014, in Chinese.
[8]Lubensky T C, Ramaswamy S and Toner J, Hydrodynamics of icosahedral quasicrystals, *Phys Rev B*, 1985, **32**, 7444-7452.

**After the publication of this paper the author and his group have obtained some solutions of some initial and boundary value problems of equations (5), (7) and (9), which examine the equations, the examination shows the equations are correct and effective, please refer to the new publications of the author and his co-workers in the field.**